\documentclass[apj]{emulateapj}
\usepackage{amsmath}
\usepackage{txfonts}
\usepackage{natbib}
\usepackage{graphicx}
\usepackage{color}
\usepackage{multirow}
\usepackage{array}
\usepackage{hyperref}
\usepackage{ulem}

\newcommand{\Msun}{M_\odot}
\newcommand{\Mbh}{M_{\mathrm{BH}}}
\newcommand{\Abh}{A_{\mathrm{BH}}}
\newcommand{\rin}{r_{\mathrm{in}}}
\newcommand{\risco}{r_{\mathrm{ISCO}}}
\newcommand{\rhori}{r_{\mathrm{BH}}}
\newcommand{\Mdotbh}{\dot{M}_{\rm BH}}

\def\ga{\,\,\raise0.14em\hbox{$>$}\kern-0.76em\lower0.28em\hbox
{$\sim$}\,\,}
\def\la{\,\,\raise0.14em\hbox{$<$}\kern-0.76em\lower0.28em\hbox
{$\sim$}\,\,}

\usepackage{soul}
\usepackage[svgnames]{xcolor}

\newcounter{maacounter}
\newenvironment{maaenvironment}[1][]{\refstepcounter{maacounter} \nobreakspace 
   {\color{Sienna}{MAA(\themaacounter):~{#1}}} \rmfamily}{}

\shorttitle{Viscous collapsar disks}

\shortauthors{Just et al.}

\begin{document}

\title{R-process viable outflows are suppressed \\ in global alpha-viscosity models of collapsar disks}

\author{O.~Just\altaffilmark{1,2}, M.~A.~Aloy\altaffilmark{3,4}, M.~Obergaulinger\altaffilmark{3},
S.~Nagataki\altaffilmark{2,5}}

\altaffiltext{1}{GSI Helmholtzzentrum f\"ur Schwerionenforschung, Planckstraße 1, D-64291 Darmstadt, Germany}
\altaffiltext{2}{Astrophysical Big Bang Laboratory, RIKEN Cluster for Pioneering Research, 2-1 Hirosawa, Wako, Saitama 351-0198, Japan}
\altaffiltext{3}{Departament d{\'{}}Astronomia i Astrof{\'i}sica, Universitat de Val{\`e}ncia, 
  Edifici d{\'{}}Investigaci{\'o} Jeroni Mu{\~n}oz, C/ Dr.~Moliner, 50, E-46100 Burjassot (Val{\`e}ncia), Spain}
\altaffiltext{4}{Observatori Astronòmic, Universitat de València, 46980 Paterna, Spain}
\altaffiltext{5}{RIKEN Interdisciplinary Theoretical and Mathematical Sciences Program (iTHEMS), 2-1 Hirosawa, Wako, Saitama 351-0198, Japan}
\email{o.just@gsi.de}

\begin{abstract}
   Collapsar disks have been proposed to be rich factories of heavy elements, but the major question of whether their outflows are neutron-rich, and could therefore represent significant sites of the rapid neutron-capture (r-) process, or dominated by iron-group elements remains unresolved. We present the first global models of collapsars that start from a stellar progenitor and self-consistently describe the evolution of the disk, its composition, and its outflows in response to the imploding stellar mantle, using energy-dependent M1 neutrino transport and an $\alpha$-viscosity to approximate turbulent angular-momentum transport. We find that a neutron-rich, neutrino-dominated accretion flow (NDAF) is established only marginally--either for short times or relatively low viscosities--because the disk tends to disintegrate into an advective disk (ADAF) already at relatively high mass-accretion rates, launching powerful outflows but preventing it from developing a hot, dense, and therefore neutron-rich core. Viscous outflows disrupt the star within $\sim$100\,s with explosion energies close to that of hypernovae. If viscosity is neglected, a stable NDAF with disk mass of about $1\,\Msun$ is formed but is unable to release neutron-rich ejecta, while it produces a relatively mild explosion powered by a neutrino-driven wind blown off its surface. With ejecta electron fractions close to 0.5, all models presumably produce large amounts of $^{56}$Ni. Our results suggest that collapsar models based on the $\alpha$-viscosity are inefficient r-process sites and that genuinely magnetohydrodynamic effects may be required to generate neutron-rich outflows. A relatively weak effective viscosity generated by magnetohydrodynamic turbulence would improve the prospects for obtaining neutron-rich ejecta.
\end{abstract}


\section{Introduction}\label{sec:introduction}

The identification of the dominant site(s) of the rapid neutron-capture (or r-) process remains a major goal of nuclear astrophysics. Neutron-star (NS) mergers are now widely considered as the most robust r-process site \citep[e.g.][]{Arnould2020f, Cowan2021g}, particularly after the first multi-messenger observation of a binary NS merger, GW170817 and its electromagnetic counterparts \citep[e.g.][]{Kasen2017a, Watson2019s}. However, whether mergers dominate galactic r-process enrichment or other sites play comparably important roles remains poorly constrained. While current models of ordinary core-collapse supernovae face serious problems in providing r-process viable conditions \citep[e.g.][]{Hudepohl2010a}, the situation for the rare class of magneto-rotational supernovae may be more optimistic. Neutron-rich outflows could be obtained in such events through the fast expansion of jets launched from a highly magnetized proto-NS, requiring however a very strong progenitor magnetic-field strength in order to enable the production of the heaviest r-process elements \citep[e.g.][]{Nishimura2015, Mosta2018q, Reichert2021t, Reichert2022f}. Another possibility, in the case of a black hole (BH) being formed (i.e. in the ``collapsar'' scenario) is that massive, subrelativistic winds get expelled from the hot and dense BH-accretion disk \citep{MacFadyen1999, Pruet2003, Kohri2005, Surman2006, Chen2007, Metzger2008b, Nagataki2007, Siegel2019b, Barnes2022o}, next to the ultrarelativistic jet that could be launched from this system and is assumed to power long gamma-ray bursts \citep[e.g.][]{Woosley1993, Popham1999, Aloy2000e, MacFadyen2001g, Komissarov2009a, Nagataki2009, Harikae2009, Tchekhovskoy2015a, Ito2015a, Aloy2021n, Gottlieb2021v}.

After several studies based on 1D models and parametrized outflow trajectories \citep[e.g.][]{Pruet2003, Surman2006} reported rather pessimistic conditions for the activation of the r-process in the relevant regime of mass-accretion rates onto the BH, $\Mdotbh$, the collapsar-disk scenario has recently seen renewed interest after \cite{Siegel2019b} pointed to the apparent similarity of these systems to remnant disks of NS mergers, for which theoretical models \citep[e.g.][]{Fernandez2013b, Just2015a, Siegel2018c, Miller2019d, Hayashi2021a}, tentatively supported by the observed kilonova of GW170817 \citep{Kasen2017a}, suggest generically neutron-rich outflows. Using general relativistic magnetohydrodynamics (MHD) simulations, \cite{Siegel2019b} assumed the ejecta produced from isolated BH disks with different initial disk masses to be representative of ejecta launched from a collapsar disk at different epochs of mass-infall rates, leading them to conclude that collapsars may be even more prolific r-process sites than NS mergers. However, apart from simplifications adopted in the neutrino treatment that may have led to an overestimated neutron richness \citep[see][for detailed discussions of neutrino-transport effects in disks]{Miller2019d, Just2021i}, the methodology of \cite{Siegel2019b} leaves open the important question of whether a disk formed during stellar collapse produces a similar outflow signature as an isolated disk for comparable values of $\Mdotbh$. In this study, we address exactly this question by conducting global simulations of collapsars, starting from a stellar progenitor model and self-consistently following the collapse, disk formation, as well as the evolution and disintegration of the disk under the influence of turbulent viscosity and neutrino-transport effects.

\begin{table}
  \centering
    \caption{\label{tab:models}Model properties and results}
    \begin{tabular}{l|cccc}
      \tableline \tableline 
                                                     & a0         & a01   & a03   & a09   \\
      \tableline      
      $\alpha$                                       & 0          & 0.01  & 0.03  & 0.09  \\
      $t_{\rm f}$ [s]                                & 200        & 100   & 100   & 100   \\
      $t_{\rm disk}$ [s]                             & 12.4       & 12.55 & 12.65 & 13.1  \\ 
      $t_{\rm NDAF}$ [s]                             & $>$\,200   & 30.6  & 14.8  & --    \\ 
      $\dot{M}_{\rm BH}(t_{\rm NDAF})$ [$M_\odot$/s] & $<$\,0.002 & 0.016 & 0.113 & --    \\ 
      $\dot{M}_{\rm ign}$ [$M_\odot$/s]              & 0          & 0.001 & 0.006 & 0.037 \\ 
      $M_{\rm BH}(t_{\rm disk})$ [$M_\odot$]         & 3.85       & 3.87  & 3.89  & 3.97  \\
      $M_{\rm BH}(t_{\rm f})$ [$M_\odot$]            & 7.14       & 5.63  & 5.27  & 4.84  \\
      $A_{\rm BH}(t_{\rm f})$                        & 0.87       & 0.81  & 0.77  & 0.71  \\
      $Y_{e,\rm min}^{\rm unb}(t_{\rm f})$           & 0.432      & 0.453 & 0.499 & 0.500 \\
      $m_{\rm unb}(t_{\rm f})$ [$M_\odot$]           & 0.82       & 7.87  & 7.98  & 8.38  \\
      $E_{\rm unb}(t_{\rm f})$ [10$^{50}$erg]        & 4.61       & 53.7  & 84.4  & 68.7  \\
      $E_{\rm jet}(t_{\rm f})$ [10$^{50}$erg]        & 437        & 108   & 25.8  & 7.20  \\
      \tableline
    \end{tabular}
    \tablecomments{From top to bottom: viscous disk parameter, final simulation time ($t_f$), disk-formation time when angular momentum at ISCO first reaches $\approx 90\,\%$ of the Keplerian value ($t_{\rm disk}$), time ($t_{\rm NDAF}$) and BH mass-accretion rate ($\dot{M}_{\rm BH}(t_{\rm NDAF})$) of NDAF-to-ADAF transition when neutrino emission efficiency drops below $1\,\%$, analytic estimate of NDAF-to-ADAF mass-accretion rate for isolated disks ($\dot{M}_{\rm ign}$; cf. Eq.~(\ref{eq:mign})), BH mass at $t_{\rm disk}$, BH mass at $t_f$, BH spin at $t_f$, minimum electron fraction within unbound material at $t_f$, mass and total (kinetic plus thermal plus gravitational) energy of gravitationally unbound material at $t_f$, total jet energy injected into the system between $t_{\rm disk}$ and $t_f$ estimated using Eq.~\eqref{eq:ljet}.}
\end{table}

%
\begin{figure*}
  \includegraphics[width=0.49\textwidth]{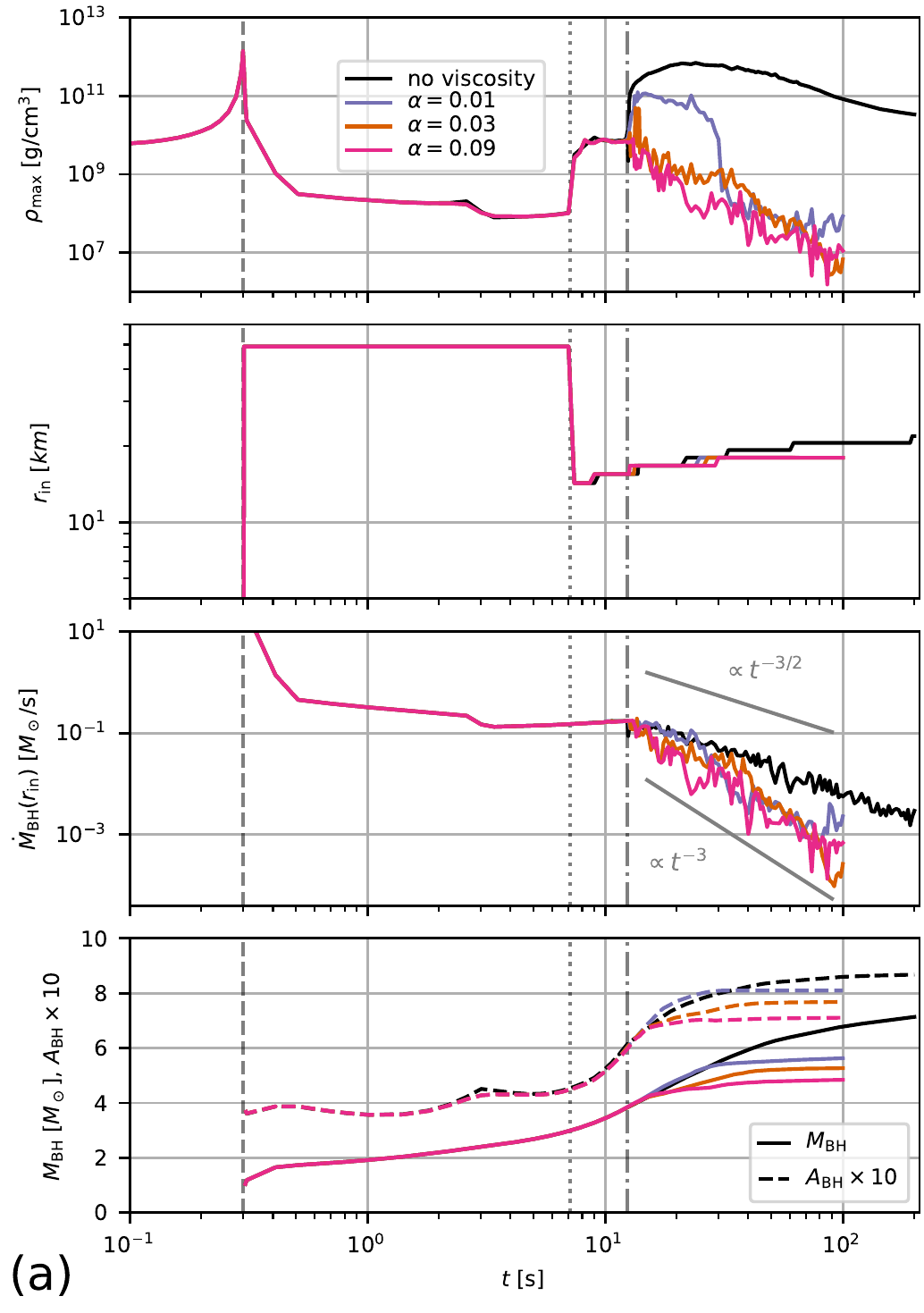}
  \includegraphics[width=0.49\textwidth]{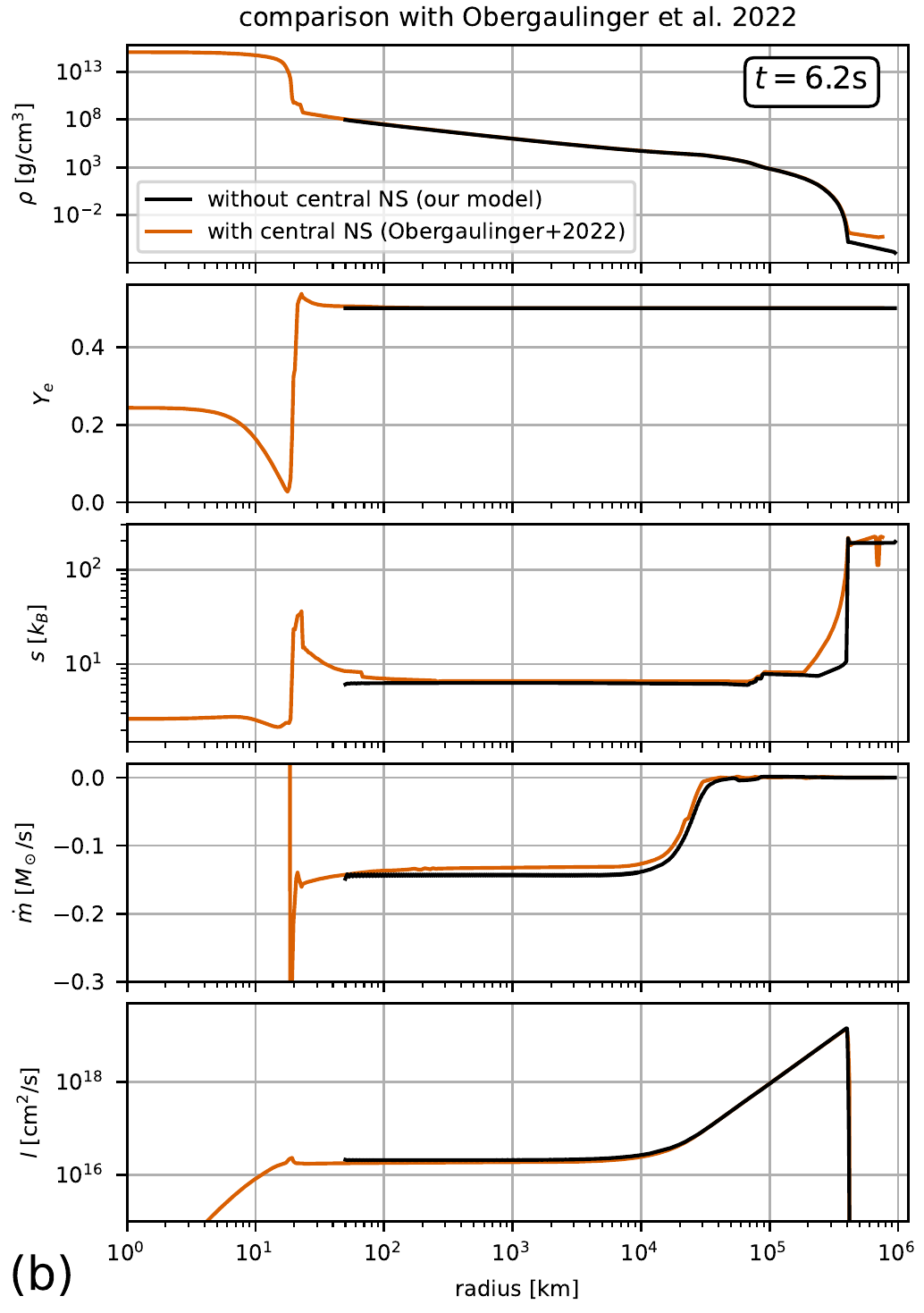}
  \caption{\emph{Left Panel:} Global properties as functions of time, namely, from top to bottom, maximum density on the computational domain, radius $\rin$ of the inner BH boundary, mass-accretion rate through the inner boundary, and mass, $\Mbh$, and spin parameter, $\Abh$, of the central BH. Time $t=0$ corresponds to the initialization of the stellar progenitor. Vertical dashed, dotted, and dot-dashed lines mark the time of core bounce, the time when $\rin$ is moved to a radius well below the ISCO, and the disk-formation time, respectively. \emph{Right Panel:} Comparison between our model (black lines), which ignores the evolution of the proto-NS, and model 16TI (orange lines; \citealp{Obergaulinger2022f}), which self-consistently follows the proto-NS evolution. From top to bottom, the density, electron fraction, entropy per baryon, radial mass flux, and specific angular momentum are depicted. 
  }
  \label{fig:martin}
\end{figure*}
\begin{figure*}
  \includegraphics[width=\textwidth]{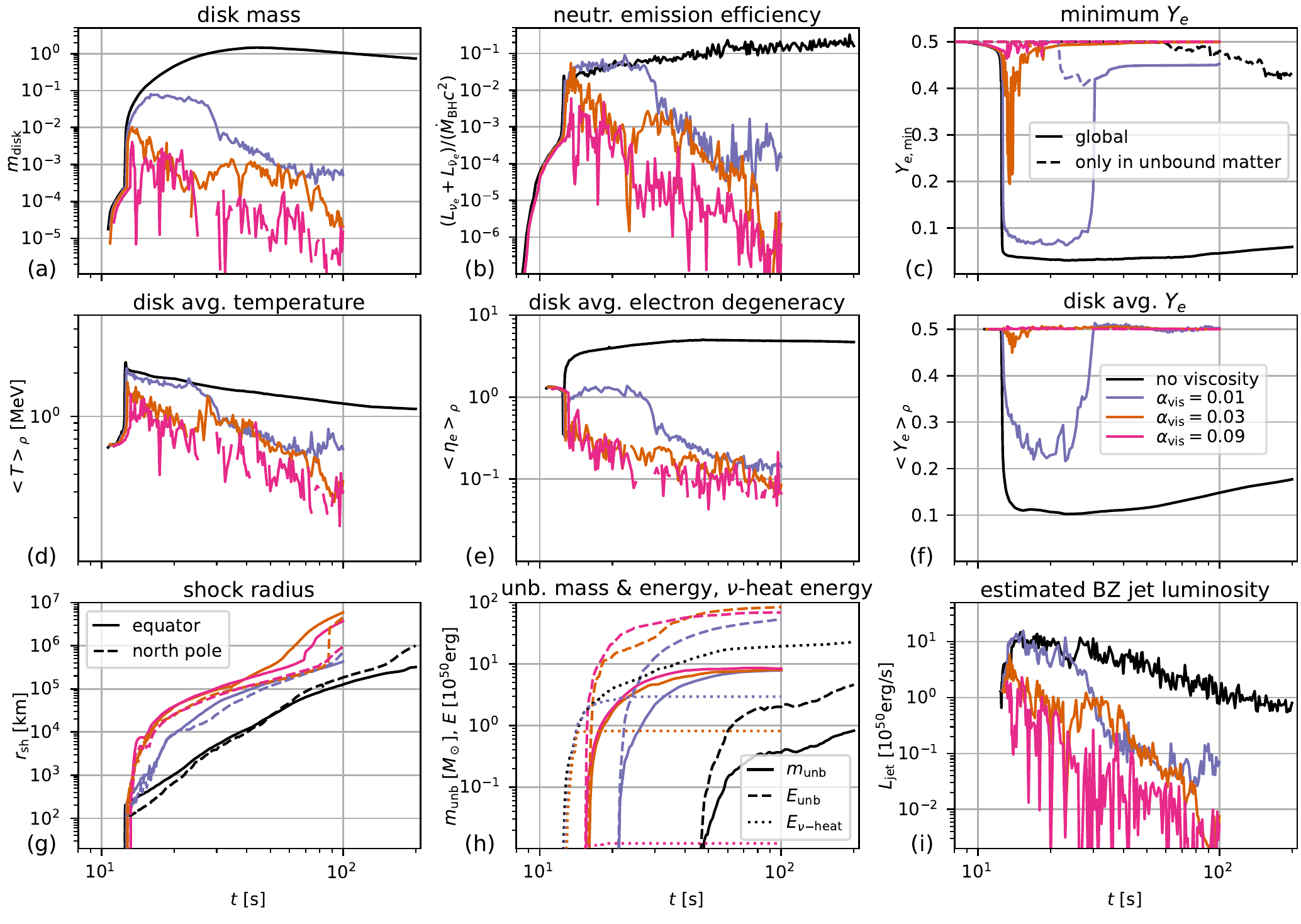}
  \caption{Global properties as functions of time characterizing the disk evolution, namely: (a) disk mass, (b) neutrino-emission efficiency (ratio of total neutrino luminosity to accreted rest-mass energy per time), (c) minimum $Y_e$ within the global domain (solid lines) and within only the gravitationally unbound material (dashed lines), (d)-(f) mass-weighted averages of temperature, electron-degeneracy parameter, and $Y_e$, respectively, over all disk material, (g) shock radii at the equator (solid lines) and north pole (dashed lines), (h) mass (solid lines) and total energy (dashed lines) instantaneously carried by all gravitationally unbound material, as well as the time- and volume-integrated net neutrino-heating rate in regions where neutrino cooling is subdominant, (i) estimated power of a jet launched via the Blandford-Znajek process (Eq.~\eqref{eq:ljet}). Missing data points for disk-related quantities indicate that no material fulfills the disk-definition criteria at these times.}
  \label{fig:glob_disk}
\end{figure*}
\begin{figure*}
  \includegraphics[width=\textwidth]{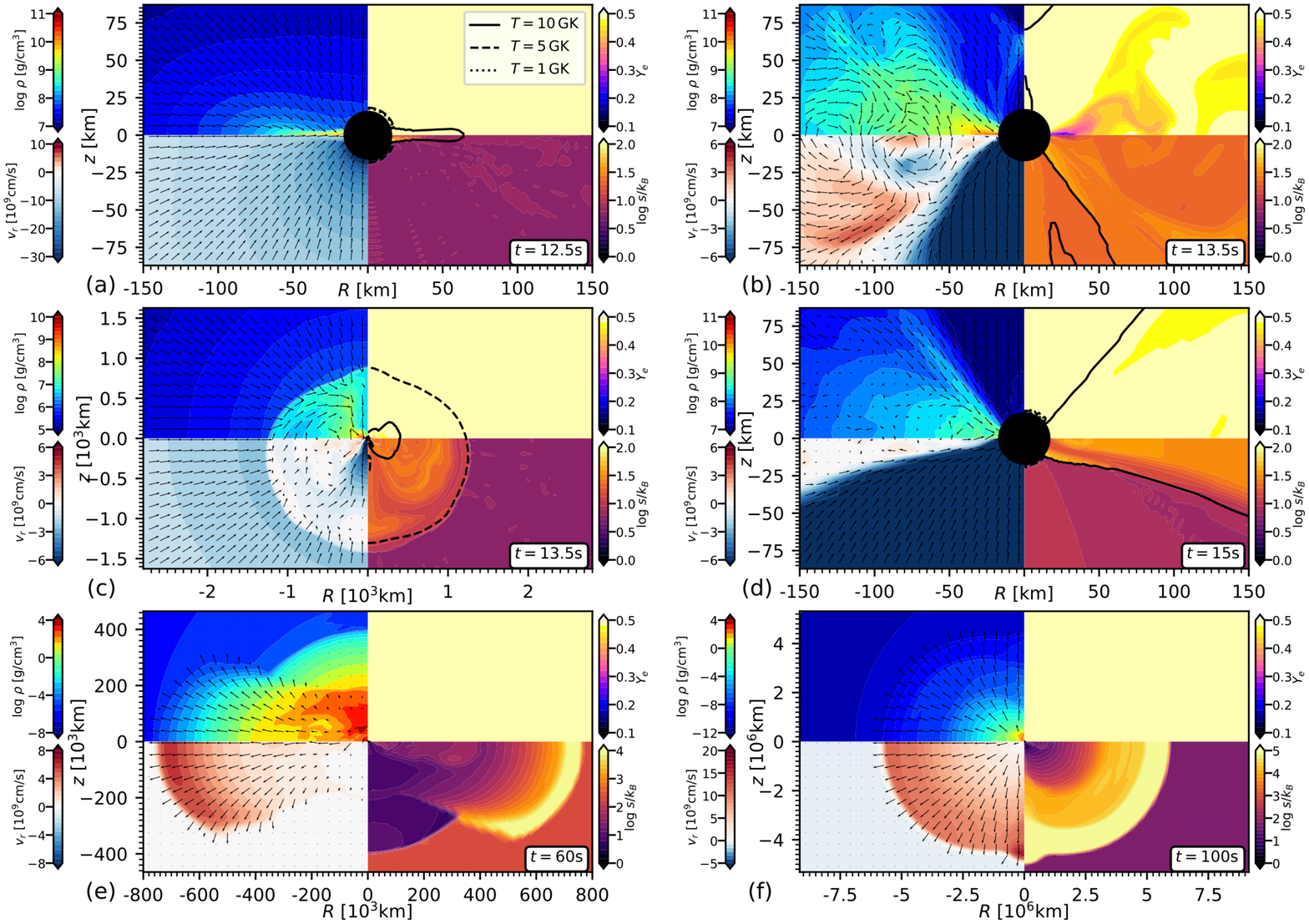}
  \caption{Color maps showing, for the viscous model a03, the distributions of density, $\rho$, radial velocity, $v_r$, electron fraction, $Y_e$, and entropy per baryon, $s/k_B$, at several characteristic times: (a) shortly before disk formation; (b) shortly after disk formation before the NDAF-to-ADAF transition; (c) for the same time as~(b) but showing the expanding shock; (d) right after the transition to ADAF; (e) shortly after shock breakout from the stellar surface; (f) during transition to the homologous expansion phase. Black arrows indicate the velocity field, while their length is capped above velocities of $10^{9}\,$cm\,s$^{-1}$. Isocontours of relevant temperatures are marked with various black lines (see legend in panel (a)).}
  \label{fig:cont_vis}
\end{figure*}
\begin{figure*}
  \includegraphics[width=\textwidth]{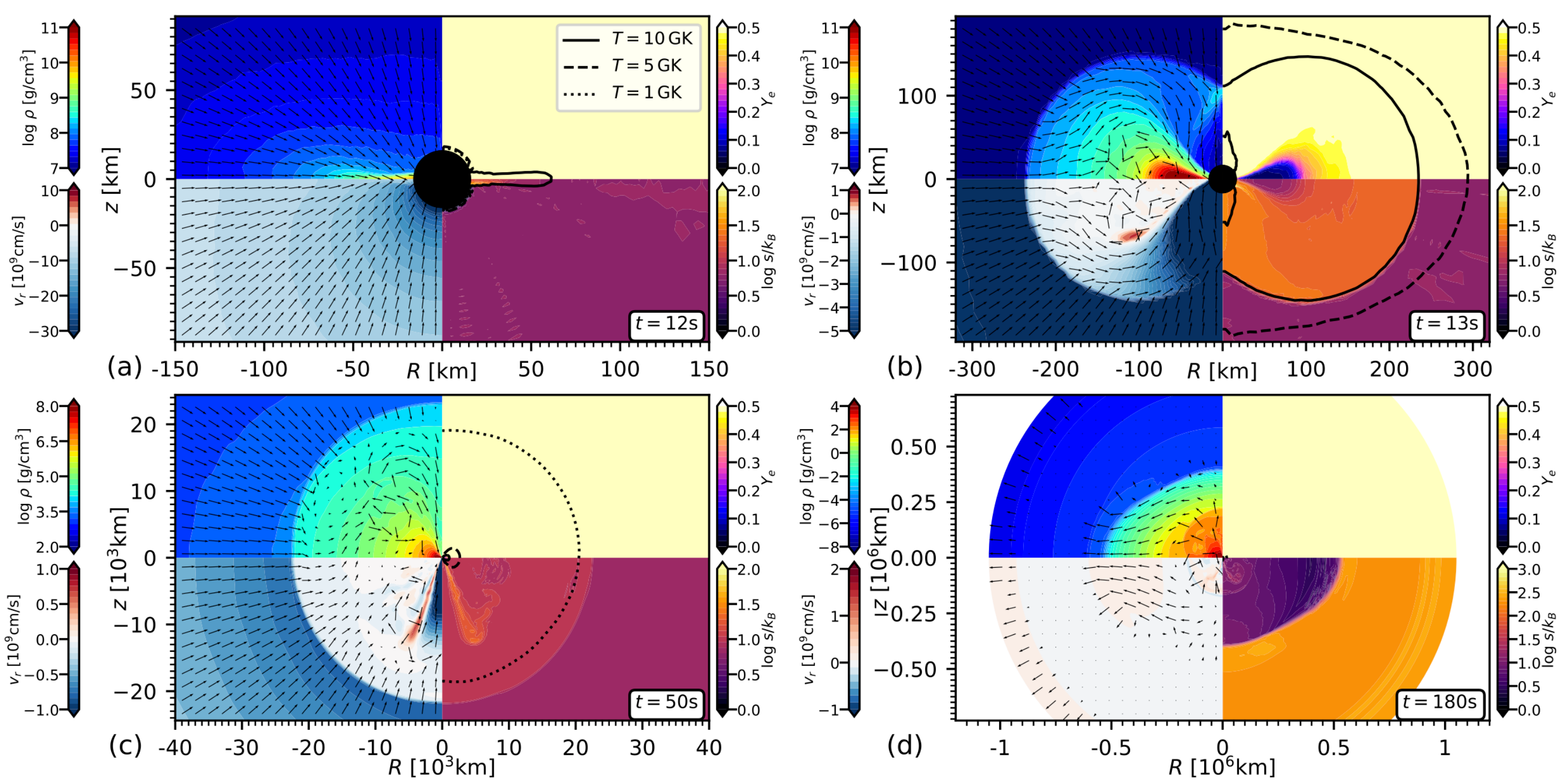}
  \caption{Same as Fig.~\ref{fig:cont_vis}, but for the non-viscous model a0, with velocity arrows whose length is capped at $10^{8}\,$cm\,s$^{-1}$, and at the characteristic times right before (a) and after disk formation (b), during the shock expansion through the stellar mantle (c), and shortly after shock breakout (d).}
  \vspace{0.6cm}
  \label{fig:cont_novis}
\end{figure*}

\section{Model setup}\label{sec:model-setup}

We adopt the strongly rotating progenitor model 16TI \citep{Woosley2006a} having a ZAMS (collapse) mass of 16 (13.96)\,$M_\odot$, which has been used in numerous existing studies of collapsars. After following the collapse until core bounce (commencing at time $t\approx 300\,$ms; cf. dashed vertical line of Fig.~\ref{fig:martin}(a)) we replace all material in the innermost region by a central BH (of mass $\Mbh$ and dimensionless spin parameter $\Abh$), which we assume would have been formed several seconds later if we had retained the central NS; this collapsar scenario is supported by the self-consistent simulations of \cite{Obergaulinger2022f}. Since for this progenitor a disk is expected to form only once material from a mass coordinate of $\sim 3.8\,M_\odot$ arrives at the innermost stable circular orbit (ISCO) \citep[cf. Fig.~2 of][]{Woosley2006a}, we initially place, to save computational resources, the absorbing inner boundary at a radius of $r_{\rm in}= 50\,$km. In order to follow the formation and evolution of the disk, we later (at about $t\sim 7\,$s) move $r_{\rm in}$ to the arithmetic average of $\risco$ and $\rhori$ (the event-horizon radius) of a Kerr BH with mass $\Mbh$ and spin $\Abh$. The parameters $\Mbh$ and $\Abh$ are continuously updated using the mass and angular momentum crossing the boundary at $\rin$, neglecting for simplicity the distinction between baryonic and gravitational mass (cf. bottom panel of Fig.~\ref{fig:martin}~(a)).

All models are simulated in axisymmetry with spherical polar coordinates using the finite-volume code ALCAR-AENUS \citep{Obergaulinger2008a,Just2015b}. We assume Newtonian physics but take into account basic general relativistic effects in the gravitational potential, computed as the sum of the BH-potential by \cite{Artemova1996} and the TOV-potential for the stellar mantle (case A of \citealp{Marek2006}). The equation of state includes neutrons, protons, helium, and $^{56}$Ni in nuclear statistical equilibrium, arbitrarily degenerate and arbitrarily relativistic electrons and positrons, and fully thermalized photons. The transport of $\nu_e$ and $\bar\nu_e$ neutrinos is treated by an energy-dependent M1 scheme \citep{Just2015b}, which takes into account the nucleonic $\beta$-processes as in \cite{Bruenn1985} augmented by weak-magnetism corrections \citep{Horowitz2002a}. Turbulent angular-momentum transport is approximated by the $\alpha$-viscosity model \citep{Shakura1973}, using the type~2 scheme of \cite{Just2015a}. In order to study the impact of viscosity, we consider three viscous cases with viscosity parameters $\alpha=0.01,\,0.03,\,$ and $0.09$ (with corresponding model names a01, a03, a09 and denoted here as low, medium, and high viscosity, respectively, motivated by comparisons with MHD disks in \citealp{Fernandez2019b, Just2021i}), and a non-viscous case ($\alpha=0$, named a0); see Table~\ref{tab:models}. The radial domain extends from $\rin$ to $r_{\rm max}$ (with $r_{\rm max}=10^{11/12}$\,cm for the non-viscous/viscous case) and is discretized in a logarithmic manner by $\approx 150$ zones per decade, while 80 uniform zones are used in the angular direction from $\theta=0$ to $\pi$. Neutrino energies are sampled by 10 bins logarithmically distributed between 0 and 80 MeV.

The models are initialized by mapping the density, $\rho$, electron fraction, $Y_e$, pressure, and angular velocity onto radial shells of the axisymmetric computational domain. We note that the outermost layers of progenitor model 16TI rotate super-Keplerian \citep{Woosley2006a} and begin to expand right after starting the simulation, however, with significantly smaller speeds than encountered later in the disk-driven outflows.

Our model setup is similar to that of \citet{MacFadyen1999} who, however, did not consistently evolve $Y_e$. In order to assess the validity of ignoring the proto-NS evolution, we compare in Fig.~\ref{fig:martin}~(a) radial profiles of our model with those taken from a model that consistently included the proto-NS (model 16TI of \citealp{Obergaulinger2022f}) at time $t=6.2\,$s, where we find good agreement.

\section{Results}\label{sec:results}

\subsection{Disk formation}\label{sec:disk-formation}

Figure~\ref{fig:glob_disk} provides additional global properties as functions of time, while Figs.~\ref{fig:cont_vis}~and~\ref{fig:cont_novis} show, for models a03 and a0, snapshots of the spatial distribution at selected times. Initially, material with specific angular momentum, $l$, lower than the corresponding Keplerian value at the ISCO, $l_{\rm ISCO,Kep}$, falls into the BH in a quasi-spherical manner. At $t\sim 13\,$s (corresponding to $\Mbh\sim 3.8\,\Msun$) gas with $l\approx l_{\rm ISCO,Kep}$ arrives at the ISCO, where it is prevented from falling into the BH by the centrifugal barrier and starts to assemble a rotationally supported disk. All models, except the one with the strongest viscosity (a09), show qualitatively the same features at the time of disk formation. Fluid elements falling in from equatorial directions accumulate at the radial outer edge of the disk, enhancing its mass (cf. Fig.~\ref{fig:glob_disk}~(a)) and leading to an immediate increase of the density (Fig.~\ref{fig:martin}~(a)),~temperature (Fig.~\ref{fig:glob_disk}~(d)),~and neutrino luminosity (Fig.~\ref{fig:glob_disk}~(b)). Since we are mainly interested in the hottest and densest regions of the flow, we define the disk here as all material within $r=500\,$km and with an angular momentum exceeding $50\,\%$ of the local Keplerian value. 
The radial deceleration due to circularization in combination with the enhanced densities and temperatures (boosting the neutrino-emission rates and leading to dissociation of nuclei into free nucleons) leaves material enough time, at least right after disk formation, to a) adapt its local $Y_e$ to the equilibrium value $Y_e^{\rm eq}$ dictated by the density, temperature and the local neutrino abundances \citep[e.g.][]{Beloborodov2003, Arcones2010, Fujibayashi2020a, Just2021i}, and b) radiate away large parts of the thermal energy produced in the disk. The system has now formed a hyper-accretion flow, or neutrino-dominated accretion flow (NDAF; e.g. \citealp{Popham1999, Kohri2005}). As is characteristic for NDAFs, the electron degeneracy (Fig.~\ref{fig:glob_disk}~(e)) close to the BH is moderately high, $\eta_e\ga 1$, resulting in a low value of $Y_e\approx Y_e^{\rm eq}\la 0.2$.

\subsection{Evolution of viscous models}\label{sec:evol-visc-models}

The subsequent evolution depends sensitively on the treatment of angular-momentum transport. We first consider the viscous models. As soon as the system enters a state of nearly Keplerian, differential rotation, viscous effects become important. In the initial NDAF state, neutrino cooling is efficient enough to approximately balance thermal heating. A necessary condition for this balance to hold is that the neutrino-cooling timescales remain shorter than or equal to the viscous heating timescales. Due to the steep temperature dependence of the former (roughly $\propto T^{-6}$), this requirement essentially translates into a condition on the disk temperature. Once the temperature drops below some critical value close to $T\sim 1\,$MeV (roughly corresponding to emission timescales of $\mathcal{O}(10\,\mathrm{s})$), viscous heating starts to dominate neutrino cooling, which triggers viscous expansion and leads to a further reduction of the temperature and the neutrino emission efficiency (Fig.~\ref{fig:glob_disk}~(b)), resulting in the freeze-out of $Y_e$ and causing the disk to undergo a transition into a radiatively inefficient disk, a so-called advection-dominated accretion flow (ADAF; \citealp{Narayan1994}). In the fiducial model with $\alpha=0.03$ this happens at about $t_{\rm NDAF}\approx 14.8\,$s (i.e. about $2\,$s after disk formation), whereas the low-viscosity model exhibits this transition at about $t\approx 30.6\,$s. In the high-viscosity model ($\alpha=0.09$) the viscosity is so strong that the disk basically never enters the NDAF phase but evolves as an ADAF right away.

The ADAF state exhibits characteristic differences to the NDAF state, in agreement with previous investigations of neutrino-cooled disks \citep[e.g.][]{Metzger2008c, Just2021i}: Once an ADAF forms, the viscously disintegrating disk is characterized by markedly lower densities and temperatures, non-degenerate electrons, and (partially) recombined nuclei. These properties are crucial to the prospects for r-process nucleosynthesis, because they imply a high equilibrium value, $Y_e^{\rm eq}\approx 0.5$ \citep[e.g.][]{Arcones2010}. In addition, due to inefficient cooling the ADAF state is subject to strong convective activity, which is why--at least in viscous but not necessarily in MHD models \citep[][]{Fernandez2019b}--outflows are launched more efficiently during the ADAF than during the NDAF phase. Finally, the NDAF-to-ADAF transition causes a steepening of $\Mdotbh$ from roughly a $t^{-3/2}$ behavior to approximately $t^{-3}$ (Fig.~\ref{fig:martin}~(a)).

As soon as the disk has formed, an accretion shock emerges as a result of the sudden deceleration of supersonically infalling stellar material by the disk environment (Fig.~\ref{fig:cont_vis}~(c)). The shock surface expands quickly (Fig.~\ref{fig:glob_disk}~(g)), powered by the energy input from the, mainly viscously driven, disk ejecta. This process ultimately leads to the unbinding of almost the entire remaining stellar mantle (Fig.~\ref{fig:glob_disk}~(h) and Table~\ref{tab:models}). The ejecta are mainly equatorial up to the point of breakout from the star (Fig.~\ref{fig:cont_vis}~(e)) and subsequently expand in lateral directions to reach an almost spherical shape at the final simulation time of $t_f=100\,$s. None of the expelled material has a particularly low $Y_e$, the minimum value (achieved in the model with the lowest viscosity) being 0.453. The total energy of the viscosity-driven explosions lies within $5-10 \times 10^{51}\,$erg, i.e. significantly higher than for ordinary core-collapse supernovae and close to that of hypernovae \citep[see, e.g.,][for a review]{Nagataki2018b}. We note that a disk-wind driven hypernova scenario was first suggested by \cite{MacFadyen1999}, but its oblate/spherical explosion geometry makes it difficult to reconcile with observations of gamma-ray bursts associated to hypernovae \citep{Mazzali2001s}, which favor a prolate and therefore more likely jet-driven supernova \citep{MacFadyen1999, MacFadyen2001g}.

\subsection{Evolution of the non-viscous model}\label{sec:evol-non-visc}

In the absence of viscous effects the disk only evolves through the effects of neutrino cooling and heating, allowing it to remain in the NDAF state for the entire simulation time of $t_f=200\,$s, with correspondingly high densities (Fig.~\ref{fig:martin}~(a)), temperatures, and electron degeneracies, as well as low values of $Y_e$ within $0.05-0.2$ (Fig.~\ref{fig:glob_disk}~(d)-(f)). Like in the viscous models, an accretion shock forms around the disk and keeps expanding, however, with significantly slower speeds because it is not fueled by viscous outflows. Correspondingly, the mass and energy of unbound ejecta is considerably lower than in the viscous models (though still increasing at $t=t_f$). The energy source powering the explosion in this model appears to be neutrino heating, which is most intense close to the BH along the surface of the disk. Exactly in this region, namely right between the main body of the disk and the polar matter inflow, a narrow outflow is launched that expands within polar angles of $\theta\sim 10-20^\circ$ (Fig.~\ref{fig:cont_novis}~(c)). The electron fraction of this neutrino-driven wind is close to $Y_e=0.5$. Support for a neutrino-driven mechanism comes from the fact that the time-integrated amount of net neutrino-heating energy (taking into account only regions where neutrino heating dominates neutrino cooling) is significantly higher than in the viscous models, high enough to explain the observed explosion energy (Fig.~\ref{fig:glob_disk}~(h)), and the neutrino efficiency is the largest among all models (Fig.~\ref{fig:glob_disk}~(b)).

\section{Discussion}\label{sec:discussion}

In contrast to $\alpha$-viscosity models of NS-merger remnant disks \citep{Fernandez2013b, Just2015a}, none of our models produces outflows with significant amounts of neutron-rich ($Y_e<0.25$) matter, even though they operate at comparable mass-accretion rates, $10^{-3}\la \dot{M}_{\rm BH}/(\Msun\,\mathrm{s}^{-1})\la 1$. Instead of r-process elements, the ejecta in our models, with $Y_e$ very close to 0.5 and (at least at early times) temperatures in excess of $5\,$GK (cf. temperature contours in Figs.~\ref{fig:cont_vis}~and~\ref{fig:cont_novis}), are likely to produce a significant amount of $^{56}$Ni (as originally suggested by \citealp{MacFadyen1999}). One characteristic difference to previously considered models of isolated disks (i.e. disks not interacting with an imploding stellar mantle) seems to be that the NDAF state is less stable and becomes disrupted at higher mass-accretion rates, as is suggested by $\dot{M}_{\rm BH}(t_{\rm NDAF})$ (cf. Table~\ref{tab:models}) being about one order of magnitude higher than the corresponding values predicted for isolated disks \citep[e.g.][]{Chen2007, De2021k, Just2021i}:
\begin{equation}\label{eq:mign}
  \dot{M}_{\rm ign} \approx 6\times 10^{-3} \left(\frac{M_{\rm BH}}{4\,M_\odot}\right)^{\frac{4}{3}}
  \left(\frac{\alpha}{0.03}\right)^{\frac{5}{3}} \, M_\odot\,\mathrm{s}^{-1}  
\end{equation}
(where we assumed $A_{\rm BH} \approx 0.8$). An additional difference to merger disks appears to be the low efficiency by which the modeled collapsar disks release the remaining neutron-rich material right around the time of freeze-out, $t_{\rm NDAF}$. While the reasons for the aforementioned differences need to be explored in future, more detailed studies, we can already conclude that they must be connected to the stellar environment into which collapsar disks are embedded, because most of the remaining properties ($M_{\rm BH}, \dot{M}_{\rm BH}, A_{\rm BH}$, and~$m_{\rm disk}$) are very similar in NS-merger disks. A critical role may be played by the radial structure of the disk: While a merger disk is born close to the ISCO and evolves as a single, expanding ring of matter, a collapsar disk is continuously replenished with matter at the circularization radius corresponding to the specific angular momentum of each infalling mass shell. Hence, even though the disk mass is similar, the radial distribution of matter, and therefore its composition, may be markedly different in both cases.

Our study contains two important shortcomings (apart from the approximate treatment of gravity): First, the $\alpha$-viscosity model employed here captures effects related to the magneto-rotational instability (MRI) only approximately. The ability to drive neutron-rich winds during the NDAF phase is expected to be higher in MHD-disks than in $\alpha$-disks \citep{Siegel2018c, Fernandez2019b}. Moreover, a full MHD treatment is necessary to reliably determine the time of the NDAF-to-ADAF transition, i.e. to assess which of our choices for $\alpha$ is most realistic. Second, we ignore any impact of the jet that is likely to be formed by the Blandford-Znajek process \citep{Blandford1977}. A jet could possibly carry away low-$Y_e$ material from the inner disk during the NDAF phase \citep{Nakamura2015b} and it would probably make the explosion geometry more prolate. A rough estimate of the jet luminosity \citep{MacFadyen2001g},
\begin{equation}\label{eq:ljet}
  L_{\rm jet} = 10^{50} \Abh^2 \left(\frac{\Mbh}{3\,\Msun}\right)^2\left(\frac{B}{10^{15}\,\rm G}\right)^2 \, \mathrm{erg\,s}^{-1},
\end{equation}
is provided in Fig.~\ref{fig:glob_disk}~(i), where we assume a magnetic-to-thermal pressure ratio of $10\,\%$ at the equatorial $\risco$, as well as flux conservation between $\risco$ and $\rhori$, to estimate the magnetic-field strength, $B$, at $\rhori$. For $\alpha>0.01$, the (crudely) estimated total energy injected by the jet between $t_{\rm disk}$ and $t_f$ (cf. Table~\ref{tab:models}) is significantly smaller than that of the viscous ejecta, and only slightly above the threshold found in \cite{Aloy2018a}, which reinforces the consistency of not modeling explicitly the jet generation for the most viscous models. Neglecting the jet feedback in low-viscosity models is not so well justified if Eq.\,\eqref{eq:ljet} is a good proxy for the jet luminosity.

Despite the aforementioned shortcomings, our study suggests that the self-consistent treatment of the imploding stellar mantle is an important ingredient for obtaining reliable predictions of $Y_e$ in collapsar disks and their outflows. Our results apply to a single potentially collapsar-forming model. Other models, where the accretion disk forms earlier and at initially higher accretion rates, could exhibit an NDAF phase for a longer time and possibly produce more neutron-rich outflows. Winds from disks at very high accretion rates are, however, expected to be subject to intense neutrino irradation, which tends to drive $Y_e$ towards 0.5 \citep{Miller2019d, Just2021i}. Future studies need to develop a more systematic understanding of the NDAF-to-ADAF transition and explore the sensitivity to the initial rotation profile and magnetic-field distribution in the stellar progenitor in order to better constrain the composition of material released by collapsar disks.

\acknowledgements  We thank the anonymous referee for constructive remarks. OJ acknowledges support by the European Research Council (ERC) under the European Union's Horizon 2020 research and innovation programme under grant agreement No. 759253. OJ and MO acknowledge support by the DFG - Project-ID 279384907 - SFB 1245. The research leading to these results has received funding from the European Union's Horizon 2020 Programme under the AHEAD2020 project (grant agreement No. 871158). MAA and MO have been supported by the Spanish Ministry of Science, Education and Universities (PGC2018-095984-B-I00 and PID2021-127495NB-I00) and the Valencian Community (PROMETEU/2019/071). MO acknowledges support from the European Research Council under grant EUROPIUM-667912, as well as from the Spanish Ministry of Science via the Ram\'on y Cajal programme (RYC2018-024938-I). SN has been supported by JSPS KAKENHI Grant Number JP19H00693 and RIKEN pioneering project "Evolution of Matter in the Universe (r-EMU)". OJ is grateful for computational support by the HOKUSAI computing facility at RIKEN and VIRGO cluster at GSI.


\end{document}